# A first approach to the Galois group of chaotic chains[1]

Stefan Groote

Loodus- ja Täppisteaduste valdkond, Füüsika Instituut,

Tartu Ülikool, W. Ostwaldi 1, 50411 Tartu, Estonia

**Abstract**

We explain in detail the definition, construction and generalisation of the Galois group of Chebyshev polynomials of high degree to the Galois group of chaotic chains. The calculations in this paper are performed for Chebyshev polynomials and chaotic chains of degree $N = 2$. Insides into possible further steps are given.

[1]Prepared for the special topic "Mathematical Modelling" of AIMS Mathematics

# 1 Introduction

Developed originally for stochastics, coupled map lattices introduced by Kaneko and Kapral [1, 2] examplify spatially extended dynamical systems with a rich structure of complex dynamical phenomena [3, 4, 5, 6, 7]. Applications for coupled map lattices can be found in models for hydrodynamical flows, turbulence, chemical reactions, biological systems, and quantum field theories (cf. Refs. [3, 4] for a review). Coupled map lattices that exhibit spatio-temporal chaotic behaviour are of particular interest. In most of the cases, the investigation of such types of coupled map lattices is mostly restricted to numerical methods, while analytical results are known only for a few exceptional cases. Among these are coupled map lattices at small coupling $a$, consisting locally of hyperbolic maps [8, 9, 10, 11] where a smooth invariant density and an ergodic behaviour can be guaranteed.

From the point of view of applications in physics, the nonhyperbolic case is certainly the more interesting one [12, 13, 14, 15, 16, 17, 18]. Applications include stochastically quantised scalar field theories [4, 19], models for vacuum fluctations and dark energy [20], and chemical kinematics [3]. For the one-dimensional case considered throughout this paper, nonhyperbolicity means that the slope of the local map is allowed to have an absolute value smaller than 1 in some regions. In this case, standard techniques from ergodic theory do not apply. However, weakly coupled Chebyshev maps of the order $N$ have been investigated analytically in a series of publications [18, 19, 20, 21, 22, 23, 24, 25, 26].

One-dimensional coupled map lattices based on Chebyshev maps with diffusive forward coupling, as considered throughout this paper, are given by the simple mathematical prescription

$$\Phi^i_{n+1} = T_N(\Phi^i_n) + \frac{a}{2}\left(T_N(\Phi^{i-1}_n) + T_N(\Phi^{i+1}_n)\right), \tag{1}$$

where $n$ is the iterative index and $i$ enumerates an (abstract) location of the fields coupled by a coupling parameter $a$. The fact that the field at the location $i$ is coupled to the fields at the nearest neighbour locations $i-1$ and $i+1$ with cyclicity assumed ($\Phi^{L+1}_n = \Phi^1_n$ for some $L \in \mathbb{N}$, $L \geq 3$) generates a one-dimensional structure which will be called a chaotic chain in the following.[2] It has been shown that a weakly coupled chain of length $L = 3$ exhibits the same behaviour as a weakly coupled longer chain [25], a fact that norishes the hope to determine the Perron–Frobenius operator of this very high-dimensional dynamical system [10, 29, 30, 31, 32] in this special case.

Besides power functions, Chebyshev maps $T_N(\Phi)$ are the only maps with the genuine property

$$T_{N^m}(T_{N^n}(\Phi)) = T_{N^{m+n}}(\Phi) \tag{2}$$

for $m, n \in \mathbb{N}$ [27]. As the uncoupled case $a = 0$ leads to the iteration $\Phi^i_{n+1} = T_N(\Phi^i_n)$, the deduced property $T_N(T_{N^n}(\Phi)) = T_{N^{n+1}}(\Phi)$ can be used to understand why Chebyshev maps are conjugate to a Bernoulli shift of $N$ symbols [28]. This conjugacy is destroyed in the coupled case where Chebyshev maps of the order $N$ have $N-1$ critical points with vanishing slope, leading to the nonhyperbolic situation. However, as Chebyshev maps are polynomials of degree $N^n$, there is some hope to get a deaper insight into the evolution of chaotic chains by considering the Galois group of these polynomials and their interations.

Restricting ourselves to the order $N = 2$, we start with the uncoupled case. In Sec. 2 we construct rational relations between the zeros of the iterative polynomials. Based on these polynomials, we explain the algorithmic composition of the Galois group. While these constructions are examples for a rigid mathematical modelling, in Sec. 3 we give an assessment on what kind of mathematical modelling will be necessary to generate a generalised Galois group for a coupled chaotic chain. In the Conclusions we summarise our results and give suggestions for further steps of mathematical modelling.

# 2 Rational relations for $N = 2$

As the Chebyshev polynomial of degree $N^n$ can be written as

$$T_{N^n}(\Phi) = \cos(N^n \arccos(\Phi)), \tag{3}$$

[2] In previous publications, the term "chaotic string" was used for this object. However, as the term "string" is used for (super)string theories, we avoid confusion by using the different and maybe even more appropriate term of chaotic chains in the following.

it is obvious that the $N^n$ zeros are equally distributed over the interval $[-\pi/2, +\pi/2]$, if the cosine (or sine) function is applied to the argument of this function. However, the composite function $T_{N^n}(\sin\theta)$ is no longer a polynomial function. Instead, we are looking at the zeros of the original Chebychev polynomial $T_{N^n}(\Phi)$. To be definite, in this section we deal with the case $N = 2$. In this case the zeros are given by nested square roots of 2,

$$\begin{aligned}
&\text{for } n=1: && \pm\frac{1}{2}\sqrt{2} \\
&\text{for } n=2: && \pm\frac{1}{2}\sqrt{2+\sqrt{2}}, \quad \pm\frac{1}{2}\sqrt{2-\sqrt{2}} \\
&\text{for } n=3: && \pm\frac{1}{2}\sqrt{2+\sqrt{2+\sqrt{2}}}, \quad \pm\frac{1}{2}\sqrt{2-\sqrt{2+\sqrt{2}}}, \\
& && \pm\frac{1}{2}\sqrt{2+\sqrt{2-\sqrt{2}}}, \quad \pm\frac{1}{2}\sqrt{2-\sqrt{2-\sqrt{2}}} \\
& && \ldots
\end{aligned} \tag{4}$$

In order to classify these square roots, we use a binary code where the digit 1 indicates the position of a minus sign in the nested square root, starting from the innermost square root. For $n = 5$ one has e.g.

$$\texttt{25 = 11001}: \qquad x_{25} = -\frac{1}{2}\sqrt{2+\sqrt{2+\sqrt{2-\sqrt{2-\sqrt{2}}}}}. \tag{5}$$

The (classical) Galois group of a polynomial is defined as the subgroup of permutations which leave invariant rational relations between the zeros of the polynomial. These relations are obtained by addition, subtraction and stepwise squaring. The simplest relations are

$$x_0 + x_1 = 0, \quad x_2 + x_3 = 0, \quad \ldots \tag{6}$$

In order to step forth, notice that in squaring a zero the outermost square root will be removed. Therefore, the sum of $x_{4i}^2$ and $x_{4i+2}^2$ is equal to 1, while the difference between these squares is the zero $x_{2i}$ of the next lowest degree. Using the third Binomial formula, this zero $x_{2i}$ can be obtained by two times the product of two zeros of the higher degree with "distance 2", i.e. $x_j^2 - x_{j+2}^2$. This procedure can be nested, and we obtain

$$n = 1: \qquad x_0 + x_1 = 0, \quad x_0^2 + x_1^2 = 1.$$

$$n=2: \quad x_0+x_1=0, \quad x_2+x_3=0,$$

$$x_0^2+x_2^2=1, \quad x_0^2-x_2^2=2x_0x_2.$$

$$n=3: \quad x_0+x_1=0, \quad x_2+x_3=0, \quad x_4+x_5=0, \quad x_6+x_7=0,$$

$$x_0^2+x_2^2=1, \quad x_0^2-x_2^2=2x_4x_6,$$

$$x_4^2+x_6^2=1, \quad x_4^2-x_6^2=2x_0x_2,$$

$$(x_0^2-x_2^2)^2+(x_4^2-x_6^2)^2=1, \quad (x_0^2-x_2^2)^2-(x_4^2-x_6^2)^2=2(x_0^2-x_2^2)(x_4^2-x_6^2).$$

$$n=4: \quad x_0+x_1=0, \quad x_2+x_3=0, \quad \ldots \quad x_{14}+x_{15}=0,$$

$$x_0^2+x_2^2=1, \quad x_0^2-x_2^2=2x_4x_6,$$

$$x_4^2+x_6^2=1, \quad x_4^2-x_6^2=2x_0x_2,$$

$$x_8^2+x_{10}^2=1, \quad x_8^2-x_{10}^2=2x_{12}x_{14},$$

$$x_{12}^2+x_{14}^2=1, \quad x_{12}^2-x_{14}^2=2x_8x_{10},$$

$$(x_0^2-x_2^2)^2+(x_4^2-x_6^2)^2=1, \quad (x_0^2-x_2^2)^2-(x_4^2-x_6^2)^2=2(x_0^2-x_2^2)(x_4^2-x_6^2),$$

$$(x_8^2-x_{10}^2)^2+(x_{12}^2-x_{14}^2)^2=1,$$

$$(x_8^2-x_{10}^2)^2-(x_{12}^2-x_{14}^2)^2=2(x_8^2-x_{10}^2)(x_{12}^2-x_{14}^2),$$

$$\left((x_0^2-x_2^2)^2-(x_4^2-x_6^2)^2\right)^2+\left((x_8^2-x_{10}^2)^2-(x_{12}^2-x_{14}^2)^2\right)^2=1,$$

$$\left((x_0^2-x_2^2)^2-(x_4^2-x_6^2)^2\right)^2-\left((x_8^2-x_{10}^2)^2-(x_{12}^2-x_{14}^2)^2\right)^2=$$

$$=2\left((x_0^2-x_2^2)^2-(x_4^2-x_6^2)^2\right)\left((x_8^2-x_{10}^2)^2-(x_{12}^2-x_{14}^2)^2\right). \tag{7}$$

Hoping that these rules are exhaustive, we can start to analyse the rules. For $n=1$, the relations does not impose any conditions on the exchange of $x_0$ and $x_1$, i.e. the Galois group is the full permutation group $P_2$. However, for $n=2$ the first line in Eq. (7) already contains a restriction, as the pairs $(x_0,x_1)$ and $(x_2,x_3)$ of independently exchangeable zeros can be only exchanged in common. While the first relation in the second line does not impose a condition on the exchange of the pairs, in the last relation this exchange will change the sign of the left hand side while the sign is kept on the right hand side. As instead of $x_0$ and $x_2$ one can use $x_1 = -x_0$ and/or $x_3 = -x_2$ as well, the exchange of the pairs $(x_0, x_1)$ and $(x_2, x_3)$ will be possible only if *either* the elements of the first pair *or* the elements of the second pair are exchanged.

## 2.1 Nestedness

As this last condition which turns out to be central for our Galois group occurs as the last condition in each of the degrees in nested form, the nestedness of the relations suggests an improvement of the notation. Denoting the two (compound) objects of the last relation by $x_0^{(n-1)}$ and $x_1^{(n-1)}$, one has

$$x_0^{(n)} := (x_0^{(n-1)})^2 - (x_1^{(n-1)})^2 = 2x_0^{(n-1)}x_1^{(n-1)}. \tag{8}$$

As mentioned before, $x_0^{(n)}$ is a root of the next lowest degree. Accordingly, this root can be identified with the pair $(x_0^{(n-1)}, x_1^{(n-1)})$, and the interchange is encoded by a sign convention,

$$\begin{aligned} x_0^{(n)} &= (x_0^{(n-1)}, x_1^{(n-1)}) \quad \text{or} \quad x_0^{(n)} = (-x_0^{(n-1)}, -x_1^{(n-1)}),\\ -x_0^{(n)} &= (x_1^{(n-1)}, -x_0^{(n-1)}) \quad \text{or} \quad -x_0^{(n)} = (-x_1^{(n-1)}, x_0^{(n-1)}), \end{aligned} \tag{9}$$

where

$$x_0^{(n-1)} := (x_0^{(n-2)})^2 - (x_1^{(n-2)})^2, \qquad x_1^{(n-1)} := (x_2^{(n-2)})^2 - (x_3^{(n-2)})^2. \tag{10}$$

For a given degree $n$ the recursion ends up with

$$x_0^{(1)} = x_0, \quad -x_0^{(1)} = x_1, \quad x_1^{(1)} = x_2, \quad -x_1^{(1)} = x_3, \quad \ldots \tag{11}$$

A general algorithm

$$\begin{aligned} x_i^{(k+1)} &= (x_{2i}^{(k)})^2 - (x_{2i+1}^{(k)})^2 \to (x_{2i}^{(k)}, x_{2i+1}^{(k)}) \quad \text{or} \quad (-x_{2i}^{(k)}, -x_{2i+1}^{(k)}),\\ -x_i^{(k+1)} &= -(x_{2i}^{(k)})^2 + (x_{2i+1}^{(k)})^2 \to (x_{2i+1}^{(k)}, -x_{2i}^{(k)}) \quad \text{or} \quad (-x_{2i+1}^{(k)}, x_{2i}^{(k)}),\\ x_i^{(1)} &\to (x_{2i}, x_{2i+1}), \quad -x_i^{(1)} \to (x_{2i+1}, x_{2i}) \end{aligned} \tag{12}$$

allows to determine the elements of the group by following the branchings by the word "or" in Eq. (12). In detail:

**n = 2:**

$$\begin{aligned}
x_0^{(2)} &\rightarrow (x_0^{(1)}, x_1^{(1)}) &\rightarrow& \;(x_0, x_1, x_2, x_3) &=& \; I \\
&\rightarrow (-x_0^{(1)}, -x_1^{(1)}) &\rightarrow& \;(x_1, x_0, x_3, x_2) &=& \; A \\
-x_0^{(2)} &\rightarrow (x_1^{(1)}, -x_0^{(1)}) &\rightarrow& \;(x_2, x_3, x_1, x_0) &=& \; B \\
&\rightarrow (-x_1^{(1)}, x_0^{(1)}) &\rightarrow& \;(x_3, x_2, x_0, x_1) &=& \; C.
\end{aligned} \tag{13}$$

The group structure is not the one of the Klein four-group containing the cyclic groups generated by the non-unit three elements as proper subgroups. Instead, this group contains only a single proper subgroup, namely the cyclic group generated by the element $A$.

**n = 3:**

$$\begin{aligned}
x_0^{(3)} \rightarrow (x_0^{(2)}, x_1^{(2)}) &\rightarrow (x_0^{(1)}, x_1^{(1)}, x_2^{(1)}, x_3^{(1)}) &\rightarrow& \;(x_0, x_1, x_2, x_3, x_4, x_5, x_6, x_7) \\
&\rightarrow (x_0^{(1)}, x_1^{(1)}, -x_2^{(1)}, -x_3^{(1)}) &\rightarrow& \;(x_0, x_1, x_2, x_3, x_5, x_4, x_7, x_6) \\
&\rightarrow (-x_0^{(1)}, -x_1^{(1)}, x_2^{(1)}, x_3^{(1)}) &\rightarrow& \;(x_1, x_0, x_3, x_2, x_4, x_5, x_6, x_7) \\
&\rightarrow (-x_0^{(1)}, -x_1^{(1)}, -x_2^{(1)}, -x_3^{(1)}) &\rightarrow& \;(x_1, x_0, x_3, x_2, x_5, x_4, x_7, x_6) \\
\rightarrow (-x_0^{(2)}, -x_1^{(2)}) &\rightarrow (x_1^{(1)}, -x_0^{(1)}, x_3^{(1)}, -x_2^{(1)}) &\rightarrow& \;(x_2, x_3, x_1, x_0, x_6, x_7, x_5, x_4) \\
&\rightarrow (x_1^{(1)}, -x_0^{(1)}, -x_3^{(1)}, x_2^{(1)}) &\rightarrow& \;(x_2, x_3, x_1, x_0, x_7, x_6, x_4, x_5) \\
&\rightarrow (-x_1^{(1)}, x_0^{(1)}, x_3^{(1)}, -x_2^{(1)}) &\rightarrow& \;(x_3, x_2, x_0, x_1, x_6, x_7, x_5, x_4) \\
&\rightarrow (-x_1^{(1)}, x_0^{(1)}, -x_3^{(1)}, x_2^{(1)}) &\rightarrow& \;(x_3, x_2, x_0, x_1, x_7, x_6, x_4, x_5) \\
-x_0^{(3)} \rightarrow (x_1^{(2)}, -x_0^{(2)}) &\rightarrow (x_2^{(1)}, x_3^{(1)}, x_1^{(1)}, -x_0^{(1)}) &\rightarrow& \;(x_4, x_5, x_6, x_7, x_2, x_3, x_1, x_0) \\
&\rightarrow (x_2^{(1)}, x_3^{(1)}, -x_1^{(1)}, x_0^{(1)}) &\rightarrow& \;(x_4, x_5, x_6, x_7, x_3, x_2, x_0, x_1) \\
&\rightarrow (-x_2^{(1)}, -x_3^{(1)}, x_1^{(1)}, -x_0^{(1)}) &\rightarrow& \;(x_5, x_4, x_7, x_6, x_2, x_3, x_1, x_0) \\
&\rightarrow (-x_2^{(1)}, -x_3^{(1)}, -x_1^{(1)}, x_0^{(1)}) &\rightarrow& \;(x_5, x_4, x_7, x_6, x_3, x_2, x_0, x_1) \\
\rightarrow (-x_1^{(2)}, x_0^{(2)}) &\rightarrow (x_3^{(1)}, -x_2^{(1)}, x_0^{(1)}, x_1^{(1)}) &\rightarrow& \;(x_6, x_7, x_5, x_4, x_0, x_1, x_2, x_3) \\
&\rightarrow (x_3^{(1)}, -x_2^{(1)}, -x_0^{(1)}, -x_1^{(1)}) &\rightarrow& \;(x_6, x_7, x_5, x_4, x_1, x_0, x_3, x_2) \\
&\rightarrow (-x_3^{(1)}, x_2^{(1)}, x_0^{(1)}, x_1^{(1)}) &\rightarrow& \;(x_7, x_6, x_4, x_5, x_0, x_1, x_2, x_3) \\
&\rightarrow (-x_3^{(1)}, x_2^{(1)}, -x_0^{(1)}, -x_1^{(1)}) &\rightarrow& \;(x_7, x_6, x_4, x_5, x_1, x_0, x_3, x_2)
\end{aligned} \tag{14}$$

## 2.2 Multiplication tables

Based on the algorithm explained before, for each degree $n$ the elements of the Galois group can be calculated and the group structure can be analysed by considering a multiplication table. For $n = 1$ we obtain the Klein-like four-group

$$V_{2,2} = \begin{array}{|c|c|c|c|} \hline I & A & B & C \\ \hline A & I & C & B \\ \hline B & C & A & I \\ \hline C & B & I & A \\ \hline \end{array} \tag{15}$$

By a systematic procedure, the subgroups of a group represented by a multiplication table can be determined. As mentioned before, the group $V_{2,2}$ has a single subgroup, namely the cyclic group $\mathbb{Z}_2$ generated by the element $A$, and this subgroup is normal. One obtains

$$\{I\} \lhd S_2 = \{I, A\} \lhd V_{2,2} = \{I, A, B, C\}. \tag{16}$$

For $n = 3$ one obtains 16 elements, and the multiplication table reads[3]

| | | | | | | | | | | | | | | | |
|---|---|---|---|---|---|---|---|---|---|---|---|---|---|---|---|
| $I$ | $A$ | $B$ | $C$ | $D$ | $E$ | $F$ | $G$ | $H$ | $J$ | $K$ | $L$ | $M$ | $N$ | $P$ | $Q$ |
| $A$ | $I$ | $C$ | $B$ | $E$ | $D$ | $G$ | $F$ | $J$ | $H$ | $L$ | $K$ | $N$ | $M$ | $Q$ | $P$ |
| $B$ | $C$ | $I$ | $A$ | $F$ | $G$ | $D$ | $E$ | $K$ | $L$ | $H$ | $J$ | $P$ | $Q$ | $M$ | $N$ |
| $C$ | $B$ | $A$ | $I$ | $G$ | $F$ | $E$ | $D$ | $L$ | $K$ | $J$ | $H$ | $Q$ | $P$ | $N$ | $M$ |
| $D$ | $E$ | $F$ | $G$ | $C$ | $B$ | $A$ | $I$ | $N$ | $M$ | $Q$ | $P$ | $K$ | $L$ | $H$ | $J$ |
| $E$ | $D$ | $G$ | $F$ | $B$ | $C$ | $I$ | $A$ | $M$ | $N$ | $P$ | $Q$ | $L$ | $K$ | $J$ | $H$ |
| $F$ | $G$ | $D$ | $E$ | $A$ | $I$ | $C$ | $B$ | $Q$ | $P$ | $N$ | $M$ | $H$ | $J$ | $K$ | $L$ |
| $G$ | $F$ | $E$ | $D$ | $I$ | $A$ | $B$ | $C$ | $P$ | $Q$ | $M$ | $N$ | $J$ | $H$ | $L$ | $K$ |
| $H$ | $K$ | $J$ | $L$ | $N$ | $Q$ | $M$ | $P$ | $D$ | $F$ | $E$ | $G$ | $A$ | $C$ | $I$ | $B$ |
| $J$ | $L$ | $H$ | $K$ | $M$ | $P$ | $N$ | $Q$ | $E$ | $G$ | $D$ | $F$ | $I$ | $B$ | $A$ | $C$ |
| $K$ | $H$ | $L$ | $J$ | $Q$ | $N$ | $P$ | $M$ | $F$ | $D$ | $G$ | $E$ | $C$ | $A$ | $B$ | $I$ |
| $L$ | $J$ | $K$ | $H$ | $P$ | $M$ | $Q$ | $N$ | $G$ | $E$ | $F$ | $D$ | $B$ | $I$ | $C$ | $A$ |
| $M$ | $P$ | $N$ | $Q$ | $K$ | $H$ | $L$ | $J$ | $B$ | $I$ | $C$ | $A$ | $D$ | $F$ | $E$ | $G$ |
| $N$ | $Q$ | $M$ | $P$ | $L$ | $J$ | $K$ | $H$ | $C$ | $A$ | $B$ | $I$ | $E$ | $G$ | $D$ | $F$ |
| $P$ | $M$ | $Q$ | $N$ | $H$ | $K$ | $J$ | $L$ | $I$ | $B$ | $A$ | $C$ | $F$ | $D$ | $G$ | $E$ |
| $Q$ | $N$ | $P$ | $M$ | $J$ | $L$ | $H$ | $K$ | $A$ | $C$ | $I$ | $B$ | $G$ | $E$ | $F$ | $D$ |

(17)

The subgroups are no longer in a one-dimensional sequence but constitute a network which is shown in Fig. 1. All subgroups are normal subgroups of the previous ones.

## 2.3 The central group $V_{2,2}$

The algorithm finds its manifestation in the group $V_{2,2}$ which is central in the Galois groups of Chebychev polynomials of degree $n \geq 1$. A representation in terms of $2 \times 2$ matrices can be found which after being derived turns out to be obvious from the construction of

[3]In order to avoid confusions, for ordinary elements I omitted the capital letters $I$ and $O$ which could be mixed up with the unit elements of the multiplicative and additive group, respectively.

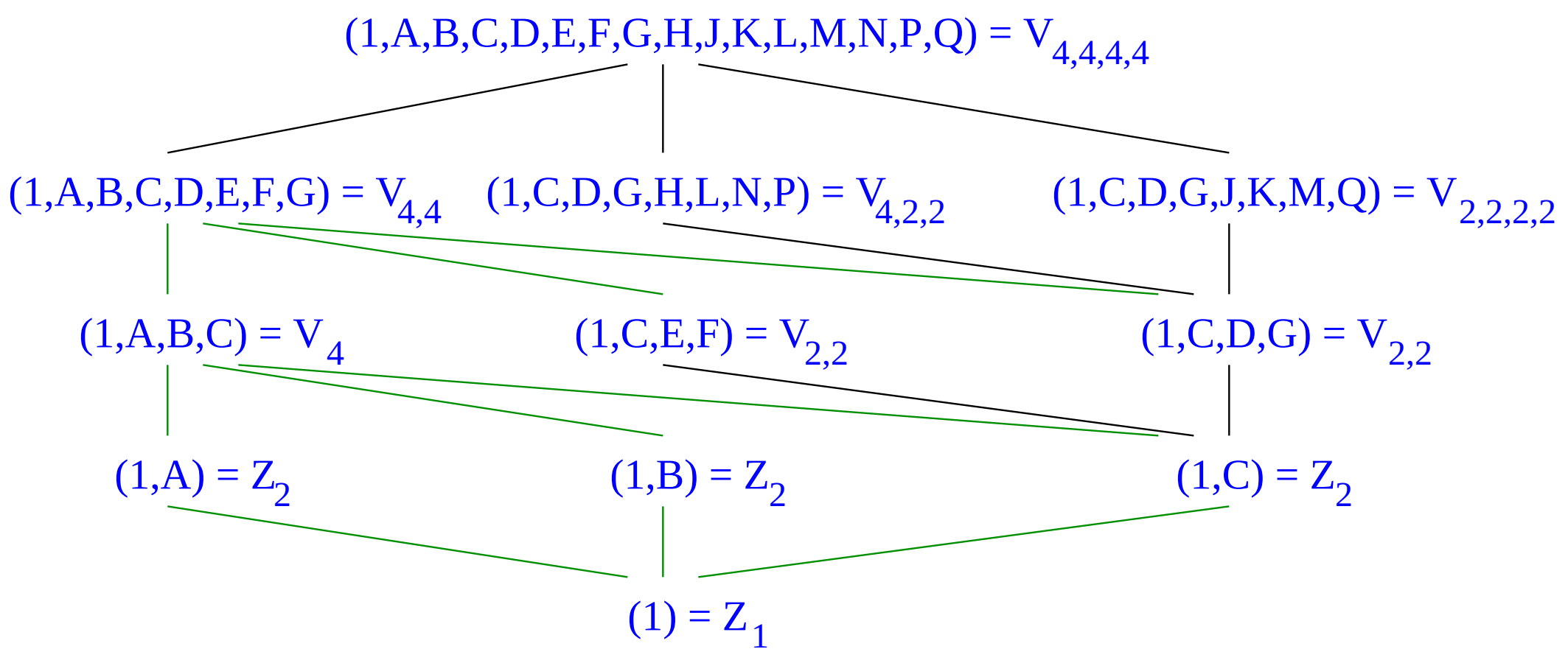

Figure 1: Galois group of $T_{2^3}(\Phi)$ and (normal) subgroups. Ideals marked in green.

the algorithm. Using the notation given for $n = 2$, one obtains[4]

$$I = \begin{pmatrix} 1 & 0 \\ 0 & 1 \end{pmatrix}, \quad A = \begin{pmatrix} -1 & 0 \\ 0 & -1 \end{pmatrix}, \quad B = \begin{pmatrix} 0 & 1 \\ -1 & 0 \end{pmatrix}, \quad C = \begin{pmatrix} 0 & -1 \\ 1 & 0 \end{pmatrix}. \tag{18}$$

As $A = -I$ and $C = -B$, both $\{I, A\}$ and $\{B, C\}$ are images of mappings of the cyclic group $\mathbb{Z}_2$. But while $\{I, A\}$ is a subgroup of the central group $V_{2,2}$, $\{B, C\}$ is a coset. Because of this, only the first one can be considered as (reducible) representation of $\mathbb{Z}_2$.

## 3 Galois group of the chaotic chain

Observables on the chaotic chain for vanishing coupling are determined by an infinitely interated Chebyshev map, and a smooth transition to an infinitely iterated map for increasing coupling is investigated in detail in Refs. [23, 24]. An insight into the behaviour of such maps can be obtained by considering iterations of high but finite degree. Such a finite degree can be taken as regularisation in the framework of a renormalisation approach, as the limit $n \to \infty$ is well understood by now and under control [26]. The zeros characterising the Galois group of the high-degree Chebyshev polynomial are traded to zeros of the highly iterated map of a coupled chaotic chain. The zeros define a partition of the interval $[-1, 1]$, and nontrivial effects on observables of chaotic chains can be observed if this partition undergoes a change of topology.

[4]Note that in case of $n = 3$, he elements $I$, $A$, $B$ and $C$ constitute the Klein four-group.

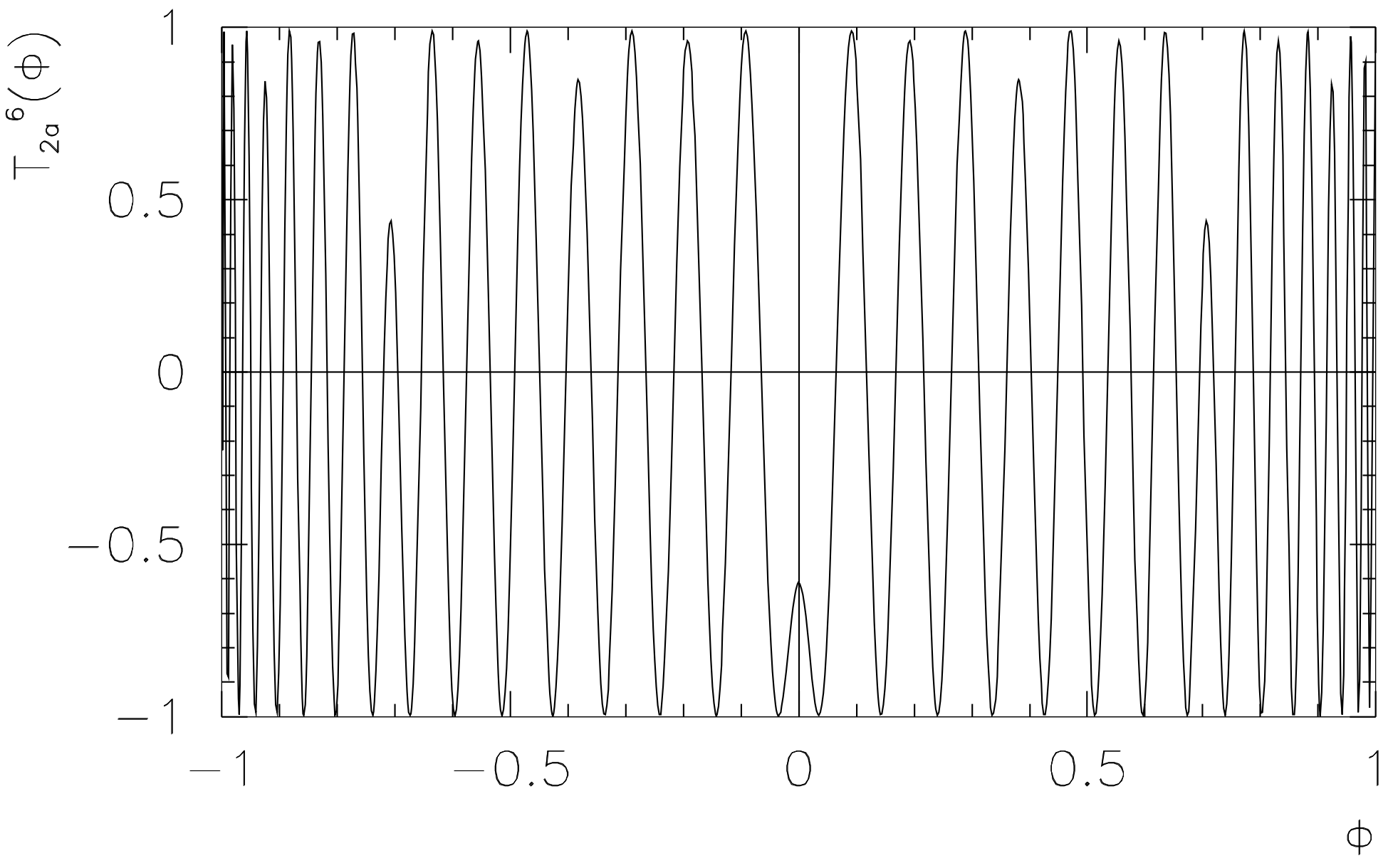

Figure 2: Iterated map for $n_{\max} = 6$ and $a = 0.00122$

### 3.1 Topology changes of the partition

If the coupling increases, starting from a Chebyshev polynomial of high degree $2^{n_{\max}}$, the iterated map of the same iterative degree $n_{\max}$ to the base $N = 2$ will be deformed smoothly. Starting with the three maxima at $\Phi = \pm 1$ and 0, sets of maxima of the map will decrease from the value $+1$ to $-1$ where they vanish between adjacent minima. Other maxima will follow, and the behaviour is approximately common to the maxima of a certain degree $n \geq 2$ located at $\sin(\pi t/2)$ with $t = \pm m/2^{n-2}$, $m = 0, 1, \ldots 2^{n-2}$. The situation is shown in Fig. 2 for the iterative degree $n_{\max} = 6$ and for the coupling $a = 0.00122$. Each time the value of such a maximum drops below zero, two zeros unite and vanish. This change induces the change of topology of the partition mentioned before.

Of course, topological changes will cause non-topological changes to the map as a whole, in particular to the zeros of the map. Effects like these are described in Ref. [26]

by the interplay of draft and blunt functions, and these effects are the reason why the behaviour is not strictly in common for maxima of the same degree. Looking at the degree $n = 4$ with eight maxima,[5] the sixteen zeros surrounding these maxima meet and vanish at different values of the coupling $a$. Using the same classification as for the zeros of the unperturbed Chebyshev polynomial in the previous section, the maxima between the zeros $x_0, x_1, x_2, x_3$ disappear at $a = 0.47 \times 10^{-3}$, between $x_8, x_9, x_{10}, x_{11}$ at $a = 0.79 \times 10^{-3}$, between $x_4, x_5, x_6, x_7$ at $a = 0.83 \times 10^{-3}$, and between $x_{12}, x_{13}, x_{14}, x_{15}$ at $a = 0.88 \times 10^{-3}$. The zeros will vanish slightly earlier (i.e. for lower values of $a$) but in the same order. In parallel to this, the structure of the Galois group will change.

## 3.2 Cascade of Galois groups

Passing these thresholds, rational relations in (7) including disappearing zeros will be skipped, leading to a smaller set of relations. In case of $n = 4$, if $x_0$, $x_1$, $x_2$ and $x_3$ disappear, the disappearence of $x_0 + x_1 = x_2 + x_3 = 0$ and $x_0^2 + x_2^2 = 1$ has no consequences on the remaining zeros. However, in skipping also $x_0^2 - x_2^2 = 2x_4x_6$ and $x_4^2 - x_6^2 = 2x_0x_2$, the zeros $x_4$, $x_5$, $x_6$ and $x_7$ will obtain an additional degree of freedom, as there are no longer restrictions to the interchange of these zeros. Therefore, the four zeros are subject to the permutative group $P_4$ while the remaining zeros $x_8, \ldots, x_{15}$ obey the relations of the degree $n = 3$. One obtains the transition $V_{4,4,4,4} \to P_4 \otimes V_{4,4}$. Passing the next threshold where $x_8$, $x_9$, $x_{10}$ and $x_{11}$ disappear, the group structure will be totally resolved to $P_4 \otimes P_4$, as these four zeros amount to half of the related zeros. Finally, at the last two thresholds the freely permutating zeros will be removed in turn. Therefore, the cascade process of degradation for the Galois group induced by the changes of the topology is given by

$$V_{4,4,4,4} \to P_4 \otimes V_{4,4} \to P_4 \otimes P_4 \to P_4 \to \mathbb{1}. \tag{19}$$

This is only the first step in analysing the topological changes in the group structure. Further considerations have to and will follow in future publications.

[5] By convention, the maxima at $\Phi = \pm 1$ are always considered as a common maximum.

### 3.3 Generalisation of the Galois group

In applying results of the previous section to chaotic chains we have lost sight of the Galois group as the subgroup of permutations of the zeros of a polynomial, leaving rational relations between these zeros invariant. Neither the zeros are related by rational relations nor the iterated coupled map is a polynomial. The only statement we can give is that both holds in the limit $a \to 0$. Due to this, a mathematical modelling for a generalisation of the Galois group applicable to chaotic chains has to use a wider understanding of the concepts of both polynomial and rationality, relating both to the definition in the limiting case $a = 0$, as we have done it intuitively in this section. The hope for obtaining such kind of generalisation is norished by the fact that in increasing the coupling, zeros will disappear but certainly not (re)appear. Therefore, this limit is always possible to perform.

## 4 Conclusions and Outlook

By keeping into touch with the uncoupled case, i.e. by performing the limit $a \to 0$ for the coupling, we have demonstrated that it is feasible to define and construct the Galois group for the chaotic chain. We have performed explicit calculations for a chaotic chain based on the Chebyshev polynomial of order $N = 2$. Similar considerations can be performed for $N = 3$. This will be subject for a future publication. In order to see the large picture of the cascade decay of degradation for the Galois group for rising coupling, a much faster code has to be written to analyse the subgroup structure for a given degree $n$. Based on this, there is hope that similar renormalisation methods as those developed in Ref. [26] can be found also for the Galois group, enabling one to perform the limit $n \to \infty$ and seeing something like a continuum limit for group and algebra. As graded $\mathbb{Z}_N$ algebras have relations to space-time metric and tetrads just in case of $N = 2$ and $N = 3$ [33, 34, 35, 36, 37, 38, 39, 40], the same cases for which chaotic chains are non-trivial, a relation of chaotic chains to space-time via generalised Galois groups is expected to exist and will be investigated in the future.

**Acknowledgements**

The author acknowledges fruitful discussions about this subject with C. Beck, R. Kerner, V. Abramov and M. Menert. The work on this subject is supported by the Estonian Science Foundation via the Centre of Excellence TK133 “Dark Side of the Universe”.